# Crystal and magnetic structure of $(La_{0.70}Ca_{0.30})(Cr_yMn_{1-y})O_3$: a neutron powder diffraction study

L. Capogna[1], A. Martinelli[2], M.G. Francesconi [3], P.G. Radaelli[4], J. Rodriguez Carvajal[5], O. Cabeza[6], M. Ferretti[2,7], C. Castellano[2], T. Corridoni[8], N. Pompeo[8]

[1] *INFM-CNR SOFT, OGG 6 Rue J. Horowitz, 38042 Grenoble – France*
[2] *INFM-LAMIA-CNR, C.so Perrone 24, 16152 Genova – Italy*
[3] *Department of Chemistry, The University of Hull, Cottingham Road, Kingston upon Hull. HU6 7RX – UK*
[4] *ISIS Facility, Rutherford Appleton Laboratory, Chilton, Didcot, Oxfordshire, OX11 0QX – UK77*
[5] *Institut Laue Langevin, 6 Rue J. Horowitz, 38042 Grenoble – France*
[6] *Dpto de Física Fac. de Ciencias, Universidade da Coruña Campus da Zapateira s/n 15071 A Coruña – Spain*
[7] *Dipartimento di Chimica e Chimica Industriale, Università di Genova, Via Dodecaneso 31, 16146 Genova – Italy*
[8] *Dipartimento di Fisica "E. Amaldi", Università degli Studi "Roma Tre", via della Vasca Navale, 84 I-00146 Roma – Italy*

**Abstract**
The crystal and magnetic structure of $(La_{0.70}Ca_{0.30})(Cr_yMn_{1-y})O_3$ for y = 0.70, 0.50 and 0.15 has been investigated using neutron powder diffraction. The three samples crystallize in the *Pnma* space group at both 290 K and 5 K and exhibit different magnetic structures at low temperature. In $(La_{0.70}Ca_{0.30})(Cr_{0.70}Mn_{0.30})O_3$, antiferromagnetic order with a propagation vector ***k*** = 0 sets in. The magnetic structure is $G_x$, i.e. of the *G*-type with spins parallel to the *a*-axis. On the basis of our Rietveld refinement and the available magnetisation data, we speculate that only $Cr^{3+}$ spins order, whereas $Mn^{4+}$ act as a random magnetic impurity. In $(La_{0.70}Ca_{0.30})(Cr_{0.50}Mn_{0.50})O_3$ the spin order is still of type $G_x$, although the net magnetic moment is smaller. No evidence for magnetic order of the Mn ions is observed. Finally, in $(La_{0.70}Ca_{0.30})(Cr_{0.15}Mn_{0.85})O_3$ a ferromagnetic ordering of the Mn spins takes place, whereas the $Cr^{3+}$ ions act as random magnetic impurities with randomly oriented spins.

## 1. Introduction

Oxides based on the different hettotype structures, produced by cations displacement or octahedral tilting from the ideal (aristotype) $ABO_3$ perovskite structure are extremely interesting materials. Their properties strongly depend on the cation located at the *B* site, which is a transition metal whereas *A* is a trivalent rare-earth. One of the most extensively studied perovskites, the calcium hole-doped manganites $(La_{1-x}Ca_x)MnO_3$ show a rich phase diagram, which is characterized by several striking structural and physical properties, such as charge, orbital and spin ordering as well as colossal magnetoresistance [1,2,3], all depending on the $[Mn^{3+}]/[Mn^{4+}]$ ratio at the *B* site.

For calcium doping concentration 0.2<x<0.5 in $(La_{1-x}Ca_x)MnO_3$, two main mechanisms are at work and in competition with each other: on the one hand the double-exchange (DE) interaction favours the $e_g$ electrons delocalization between $Mn^{3+}$ and $Mn^{4+}$ ions in the presence of a ferromagnetic (FM) order of their $t_{2g}$ spin cores. On the other hand a strong electron-phonon coupling arising from the Jahn-Teller (JT) distortion of the $Mn^{3+}$-$O_6$ octahedra [4] tend to localize the electrons.

Substituting Cr for Mn in $(La_{1-x}Ca_x)MnO_3$ could modulate this competition [5,6]. It is indeed expected that Cr, by assuming the form of $Cr^{3+}$, hinders the JT distortions, thanks to its external electronic configuration $t_{2g}^3$ which is the same as that of Mn$^{4+}$. Moreover, one would expect a participation of Cr$^{3+}$ to DE, as suggested by the effect of Cr substitution in Pr$_{0.5}$Ca$_{0.5}$MnO$_3$, which destroys the AFM, the charge and the orbital ordering in the Mn sub-lattice, favouring ferromagnetism [7].

However, macroscopic investigations based on magnetization and resistivity measurements on (La$_{0.70}$Ca$_{0.30}$)(Cr$_y$Mn$_{1-y}$)O$_3$ compounds have suggested that Cr$^{3+}$, despite its external electronic configuration, is not involved in the DE [5,6].

In this paper we tackle these issues from a microscopic point of view by studying the nuclear and the magnetic structures of (La$_{0.70}$Ca$_{0.30}$)(Cr$_y$Mn$_{1-y}$)O$_3$ (y=0.70, 0.50, 0.15) with neutron powder diffraction.

## 2. Experimental

The (La$_{0.70}$Ca$_{0.30}$)(Cr$_y$Mn$_{1-y}$)O$_3$ samples (y = 0.70, 0.50 and 0.15) were prepared by a standard solid state reaction in air. Stoichiometric amounts of MnO$_2$, La$_2$O$_3$, CaCO$_3$ and Cr$_2$O$_3$ were intimately mixed together, pelletized and heated at 1473 K for 12 hours. The mixtures were then re-ground and re-heated at 1543 K for further 24 hours. The resulting products were characterized by X-ray powder diffraction (Cu K$\alpha$ radiation, Siemens D5000).

Magnetization measurements were carried using a vibrating sample magnetometer; details are reported in Ref. [5, 6].

Neutron powder diffraction (NPD) data were collected on the high resolution powder diffractometer D2B at the Institut Laue Langevin, Grenoble using a wavelength $\lambda$ = 1.5940 Å. A standard orange cryostat was used to lower the temperature to 5 K, well below the magnetic transition. For each sample, full diffraction patterns (2θ range 5°-160°) were collected at 290 K and 5 K. The crystal and magnetic structures were then refined with the Rietveld method [8] using the program FULLPROF [9].

## 3. Results and discussion

In (La$_{0.70}$Ca$_{0.30}$)MnO$_3$, charge balance imposes a nominal [Mn$^{3+}$]/[Mn$^{4+}$] ratio equal to 70/30. Upon progressive substitution of Cr$^{3+}$ for Mn, the Mn$^{3+}$ content decreases. The [Mn$^{3+}$]/[Mn$^{4+}$] ratio becomes 65/35 and 40/60 (for y = 0.15 and 0.50, respectively), whereas for y = 0.70 there is no more mixed valence at the B site since all Mn ions are tetravalent. The results of the crystal and magnetic structure refinement are reported for all samples in Tables 1 and 2 at 290 K and 5 K, respectively.

|  |  |  | y =0.70 | y =0.50 | y =0.15 |
|---|---|---|---|---|---|
|  | a (Å) |  | 5.4319(1) | 5.4390(2) | 5.4571(9) |
|  | b (Å) |  | 7.6847(2) | 7.6912(3) | 7.7162(2) |
|  | c (Å) |  | 5.4552(2) | 5.4596(3) | 5.472(1) |
| La/Ca | 4c | x | 0.0162(4) | 0.0188(3) | 0.0221(7) |
|  |  | y | – | – | – |
|  |  | z | -0.0031(5) | -0.0046(5) | -0.0087(9) |
| Cr/Mn | 4b | x | 0 | 0 | 0 |
|  |  | y | 0 | 0 | 0 |
|  |  | z | – | – | – |
| O$_{ax}$ | 4c | x | 0.4923(4) | 0.4921(1) | 0.4939(1) |
|  |  | y | – | – | – |
|  |  | z | 0.0607(4) | 0.0617(1) | 0.0655(1) |
| O$_{eq}$ | 8d | x | 0.2756(3) | 0.2753(1) | 0.2752(1) |
|  |  | y | 0.0318(2) | 0.0316(1) | 0.0298(1) |
|  |  | z | 0.7258(3) | 0.7260(1) | 0.7216(1) |
|  | R$_{Bragg}$ (%) |  | 5.51 | 3.68 | 4.84 |
|  | R$_F$ (%) |  | 4.95 | 2.73 | 3.52 |

**Table 1**: Unit cell dimensions, atomic positional parameters and R factors of (La$_{0.70}$Ca$_{0.30}$)(Cr$_y$Mn$_{1-y}$)O$_3$ at 290 K from Rietveld refinements.

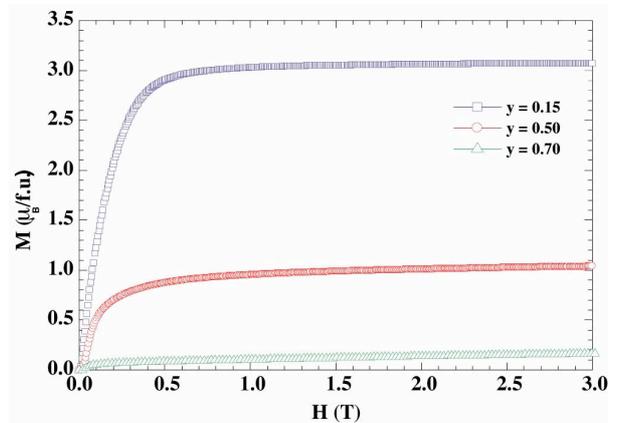

**Figure 1**: Magnetization as a function of an applied magnetic field in (La$_{0.70}$Ca$_{0.30}$)(Cr$_y$Mn$_{1-y}$)O$_3$ for y = 0.15, 0.50 and 0.70 (data collected at 10 K).

We observe that the unit cell parameters at both temperatures show a slightly decreasing trend as the Cr content increases. This is consistent with $Cr^{3+}$ substituting for $Mn^{3+}$ only as the latter is greater than $Mn^{4+}$ (ionic radii are:

$IR_{Cr^{3+}}$ = 0.615 Å;

$IR_{Mn^{3+}}$ (HS) = 0.645 Å;

$IR_{Mn^{4+}}$ (HS) = 0.530 Å [10]).

Figure 1 shows the magnetization of the samples at 10 K, as a function of the applied magnetic field. As observed in previous studies [5,6], for $y$ = 0.70 and $y$ = 0.50 saturation is not achieved even at 12 T, where magnetizations are ~ 0.15 $\mu_B$ and ~ 1.0 $\mu_B$ respectively, while for $y$ = 0.15 a saturation magnetization of ~ 3.10 $\mu_B$, reached at 3 T, can be assumed. The rising of the saturation field and the decreasing of saturation magnetization with Cr amount suggest that Cr substitution generally hinders FM behaviour in $(La_{0.70}Ca_{0.30})(Cr_yMn_{1-y})O_3$.

|  |  |  | $y$ =0.70 | $y$ =0.50 | $y$ =0.15 |
|---|---|---|---|---|---|
|  | $a$ (Å) |  | 5.4244(1) | 5.4324(2) | 5.4497(9) |
|  | $b$ (Å) |  | 7.6722(2) | 7.6808(3) | 7.6995(9) |
|  | $c$ (Å) |  | 5.4455(1) | 5.4505(2) | 5.4631(9) |
| **La/Ca** | 4$c$ | $x$ | 0.0192(3) | 0.0211(1) | 0.0221(7) |
|  |  | $y$ | – | – | – |
|  |  | $z$ | -0.0048(4) | -0.0054(1) | -0.0080(9) |
| **Cr/Mn** | 4$b$ | $x$ | 0 | 0 | 0 |
|  |  | $y$ | 0 | 0 | 0 |
|  |  | $z$ | – | – | – |
| **O$_{ax}$** | 4$c$ | $x$ | 0.4921(4) | 0.4925(1) | 0.4943(1) |
|  |  | $y$ | – | – | – |
|  |  | $z$ | 0.0607(3) | 0.0633(1) | 0.0598(1) |
| **O$_{eq}$** | 8$d$ | $x$ | 0.2768(2) | 0.2763(1) | 0.2745(1) |
|  |  | $y$ | 0.0327(2) | 0.0321(1) | 0.0331(1) |
|  |  | $z$ | 0.7253(2) | 0.7254(1) | 0.72048(1) |
|  | $R_{Bragg}$ (%) |  | 4.96 | 3.99 | 6.36 |
|  | $R_F$ (%) |  | 3.95 | 2.71 | 4.73 |

**Table 2:** Unit cell dimensions, atomic positional parameters and $R$ factors of $(La_{0.70}Ca_{0.30})(Cr_yMn_{1-y})O_3$ at 5 K from Rietveld refinements.

### 3.1 $(La_{0.70}Ca_{0.30})(Cr_{0.70}Mn_{0.30})O_3$

The comparison of the NPD data collected at 290 K and 5 K shows the onset of an AFM ordering at low temperature. At the same time, the background is significantly reduced because the incoherent scattering associated with the spin disorder in the PM phase is suppressed. Figure 2 shows the Rietveld refinement plot of the data collected at 5 K; at this temperature a notable contraction of the cell edges is observed. This structural change does not involve a rearrangement of the octahedral tilting, since the O-$B$-O bond angles are almost coincident with those obtained at 290 K.

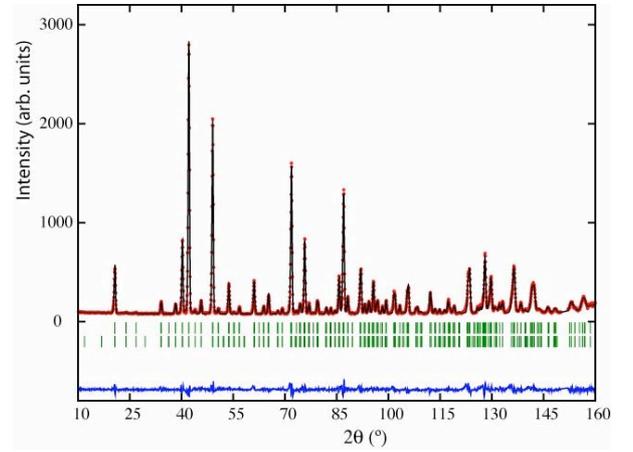

**Figure 2**: Rietveld refinement plot of the NPD data of $(La_{0.70}Ca_{0.30})(Cr_{0.70}Mn_{0.30})O_3$ collected at 5K on the D2b diffractometer (ILL) with λ = 1.5940 Å. The red points represent the experimental values and the solid black line the calculated profile. The upper marks describe the nuclear reflections while the lower set indicates the magnetic ones. The crystal space is *Pnma* and the magnetic structure is antiferromagnetic of type $G_x$.

$(La_{0.70}Ca_{0.30})(Cr_{0.70}Mn_{0.30})O_3$ crystallizes in the *Pnma* space group at room temperature and retains this structure also at low $T$. A significant improvement of the fit is obtained by refining the strain parameters. Strain is probably related to the different ionic radii of $Cr^{3+}$ and $Mn^{4+}$ (0.615 Å and 0.530 Å, respectively [10]), both located at the $B$ site. At 290 K the $B$ atom is located at the centre of an almost undistorted octahedron characterized by three very similar $B$-O bond lengths, whereas at low $T$ a slight distortion is observed which is probably related to the contraction of the cell edges.

The AFM diffraction peaks observed in the NPD pattern at 5 K can be indexed with a propagation vector $k=0$. Irreducible representations analysis [11] yields two possible configurations that are compatible with the experimental AFM $G$-type structure: $\Gamma_1$ ($G_x$, $C_y$, $A_z$) and $\Gamma_5$ ($A_x$, $F_y$, $G_z$). An accurate analysis of the NPD data reveals that the magnetic structure belongs to the $Pnma$ Shubnikov space group with the magnetic moments aligned along $x$ ($\Gamma_1$). Different magnetic structural models have been tested in order to ascertain if the spins of $Cr^{3+}$ are somehow ordered with those of $Mn^{4+}$, that is if magnetic interactions take place between $Cr^{3+}$ and $Mn^{4+}$ cations through the intervening O.

Assuming that both ion species take part in the magnetic interactions, we obtain a magnetic moment of 2.22(2) $\mu_B$ and $R_{Magneticc}$ factor is 2.43%. The refined magnetic moment at the $B$ site is ~ 70% of the theoretical spin only value (3.0 $\mu_B$, from the external electronic configuration $t_{2g}^3$). This would suggest that only 70% of the magnetic cations are magnetically ordered, a value that matches the amount of substituting $Cr^{3+}$. Indeed, the best fit result ($R_{Magnetic}$=1.99%) is obtained assuming that $Cr^{3+}$ are the only cations magnetically ordered. In this case the ordered magnetic moment of $Cr^{3+}$ is equal to 3.20(2) $\mu_B$, consistent with the expected theoretical spin only value. Conversely, if only $Mn^{4+}$ are considered with an occupation of 70%, the agreement is somewhat worse ($R_{Magnetic}$=3.91%). This difference in the R factors reflect the fact that the magnetic form factors of $Cr^{3+}$ and $Mn^{4+}$ are somehow different at wavevectors Q of magnetic interest. This difference does not per se allow any firm conclusion to be drawn on which ionic species takes part in the magnetic order. However the available magnetisation data [5] suggests clearly that superexchange interactions occur only among neighbouring $Cr^{3+}$ cations through the intervening $O^{2-}$.

We can then speculate that in $(La_{0.70}Ca_{0.30})(Cr_{0.70}Mn_{0.30})O_3$ magnetic interactions between neighbouring $Cr^{3+}$ and $Mn^{4+}$ are hindered, although both cations share the same external electronic configuration $t_{2g}^3$, and that $Mn^{4+}$ act as random magnetic impurity in the $B$ sub-lattice.

The observed magnetic structure is consistent with that detected in $(La_{1-x}Ca_x)CrO_3$ compounds ($x \leq 0.30$). These materials, which are potentially useful for application in solid oxide fuel cells [12], exhibit a $G_x$-type spin ordering coupled with a $A_z$-type one. The magnetic moments increase with the content of calcium, due to the associated increase of $Cr^{4+}$. Since in $(La_{0.70}Ca_{0.30})(Cr_{0.70}Mn_{0.30})O_3$ the whole Cr is in the trivalent state, the $A_z$ component is hindered, as concluded for LaCrO3 [12] where the $A_z$ contribution is extremely faint.

This result is quite intriguing and deserved to be discussed in detail. It is well known that different configurations can contribute to superexchange interaction [13,14]; in particular the cation-cation transfer configurations may be more important than others and the overlap integral between the two neighbouring cation $d$ orbitals contributes significantly to the exchange integral [13]. In this context a phenomenological law has been proposed which relates superexchange with the metal-oxygen bond length [13]. In CaMnO3 all the manganese ions are $Mn^{4+}$ and their spins order at low $T$ with a G-type magnetic lattice [15]. The same magnetic structure is observed in LaCrO3, where all the chromium ions are $Cr^{3+}$ [12]. In CaMnO3 the $Mn^{4+}$-O bonds do not exceed 1.90 Å [15], whereas in LaCrO3 the $Cr^{3+}$-O bond lengths is about 1.95 Å, a value matching that observed in $(La_{0.70}Ca_{0.30})(Cr_{0.70}Mn_{0.30})O_3$. Consequently the elongation of the $Mn^{4+}$-O bond coupled with the reduced size of $Mn^{4+}$ (compared to $Cr^{3+}$) strongly reduces the overlap between the $d$ orbitals of neighbouring $Mn^{4+}$ and $Cr^{3+}$ as well as the superexchange interaction between $Mn^{4+}$ and O. As a result it can be expected that in the $B$ sub-lattice a $Cr^{3+}$ cation will preferentially interact with neighbouring $Cr^{3+}$ and that the $Cr^{3+}$-$Cr^{3+}$ coupling is favoured over the $Cr^{3+}$-$Mn^{4+}$ one.

The small magnetization measured in saturating conditions is thus consistent with the fact that the AFM ordering of the Cr sub-lattice is not destroyed by the applied magnetic field, whereas the Mn sub-lattice can FM order locally.

### 3.2 $(La_{0.70}Ca_{0.30})(Cr_{0.50}Mn_{0.50})O_3$

The neutron diffraction pattern (Figure 3) of $(La_{0.70}Ca_{0.30})(Cr_{0.50}Mn_{0.50})O_3$ collected at 290 K shows an anisotropic asymmetric line broadening of the diffraction peaks. The single nominal phase yields a reasonable fit. However, a notable improvement of the $R$ factors and fitting profile is obtained assuming the coexistence of two different isostructural phases characterized by slightly different [Mn]/[Cr] ratios. The inhomogeneous distribution of Mn and Cr could be due to an incipient phase separation, occurred during the cooling stage of the sample preparation, or to an incompleteness of the reaction. According to our Rietveld refinement, the main phase constitutes about 86% of the total molar content and is characterized by a [Mn]/[Cr] ratio equal to ~ 1.20, a value quite close to the nominal one (1.00). The lattice and structural parameters of this phase are reported in Tables 1 and 2.

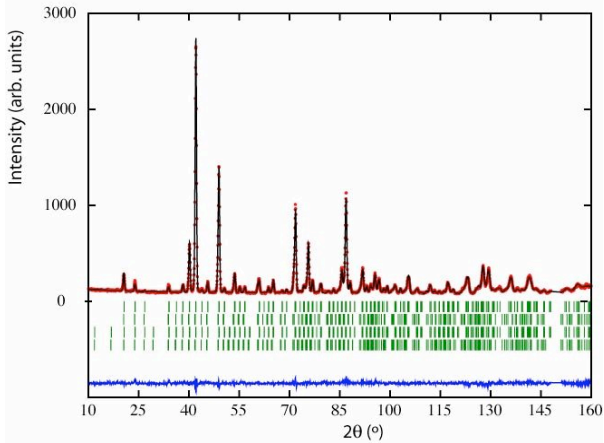

**Figure 3**: Rietveld refinement plot of the NPD data of $(La_{0.70}Ca_{0.30})(Cr_{0.50}Mn_{0.50})O_3$ collected at 5K on the D2b diffractometer (ILL) with $\lambda$ = 1.5940 Å. The experimental data (red points) are better described by two isostructural phases (upper two sets of marks). The two lower sets are the magnetic reflections of the two nuclear phases. The crystal space is *Pnma* and the magnetic structures are antiferromagnetic of type $G_x$.

The secondary phase is probably better described by a mixture of Cr-enriched isostructural phases exhibiting slight variations of the [Mn]/[Cr] ratio. In any case this sample gives a clear indication of how Mn and Cr behave when they share the same atomic site with about the same degree of occupancy. The NPD pattern at 5 K shows the same AFM peaks ($G$-type) observed in the sample ($La_{0.70}Ca_{0.30})(Cr_{0.70}Mn_{0.30})O_3$, although their intensities are decreased (Figure 5). Since Cr is known to destroy AFM ordering in the Mn sub-lattice of manganites, this suggests that this peak arise from an AFM structure occurring in the Cr sub-lattice. Note that the peak at $2\theta \sim 24°$ in Figure 5 is due to the nuclear structure for $y$ = 0.50 (it is observed also in the NPD pattern collected at 290 K), whereas has a magnetic origin in the sample with $y$ = 0.15.

Small angle neutron scattering (SANS) is essentially the same at 290 K and 5 K. In manganites SANS contribution is related to the development of FM clusters in the PM region [16] and the onset of a long-range magnetic order suppresses this contribution. In $(La_{0.70}Ca_{0.30})(Cr_{0.50}Mn_{0.50})O_3$ the presence of SANS at low $T$ is probably related to the fact that long-range magnetic order does not take place in the Mn sub-lattice at low $T$ and hence FM clusters do not percolate on cooling. A similar behaviour characterizes also the $(La_{0.70}Ca_{0.30})(Cr_{0.70}Mn_{0.30})O_3$ sample, although in this case it is partially masked by the evolution of the incoherent scattering with $T$. Contrary, in $(La_{0.70}Ca_{0.30})(Cr_{0.50}Mn_{0.50})O_3$ the incoherent scattering exhibits a faint decrease with $T$ because of it is dominated not by spin disorder as in the previous case, but by chemical disorder at the $B$ site.

The magnetization under saturation of this sample is consistent with the scenario already proposed for $(La_{0.70}Ca_{0.30})(Cr_{0.70}Mn_{0.30})O_3$, where the applied magnetic field does not destroy the AFM ordering of the Cr sub-lattice, but aligns ferromagnetically a fraction of the Mn spins.

In the Rietveld refinement, several models have been tested for the magnetic structure. In particular, the addition of a ferromagnetic component does not improve the R factors and yields a FM moment which is too small to have physical significance. We therefore conclude that this Cr concentration is characterized by the same magnetic structure as (La$_{0.70}$Ca$_{0.30}$)(Cr$_{0.70}$Mn$_{0.30}$)O$_3$ with an AFM moment on the Cr ions equal to 2.40(1) $\mu_B$.

### 3.3 (La$_{0.70}$Ca$_{0.30}$)(Cr$_{0.15}$Mn$_{0.85}$)O$_3$

For this chromium concentration, saturation is reached at 3 T and the magnetisation is ~ 3.10 $\mu_B$; this value matches exactly what is expected considering only the Mn sub-lattice FM ordered. If both Mn and Cr cations are considered FM aligned the resulting nominal moment is 3.55 $\mu_B$, a value which exceeds by far the experimental one. This suggests that Cr$^{3+}$ does not participate in the mechanism of double exchange that takes place between the Mn$^{3+}$ and Mn$^{4+}$ cations and leads to the onset of the FM spin order. We have once again tested several magnetic models using the Rietveld method in order to determine what cationic species are FM magnetically ordered. The best agreement is obtained assuming that only the Mn ions are FM ordered along the $x$ axis, whereas Cr$^{3+}$ ions act as random impurities. The value of the Mn spin is found to be 2.90(4) $\mu_B$.

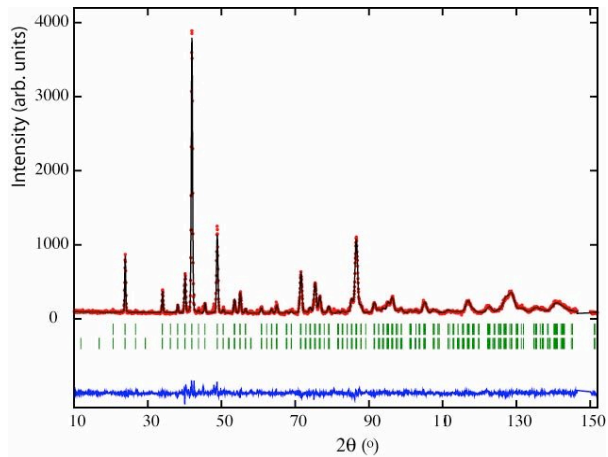

**Figure 4**: Rietveld refinement plot of the NPD data of (La$_{0.70}$Ca$_{0.30}$)(Cr$_{0.15}$Mn$_{0.85}$)O$_3$ collected at 5K on the D2b diffractometer (ILL) with $\lambda$ = 1.5940 Å. The red points represent the experimental values and the solid black line the calculated profile. The upper marks describe the nuclear reflections while the lower set corresponds to the magnetic ones. The crystal space is *Pnma* and the magnetic structure is ferromagnetic.

Our analysis gives strong microscopic evidence of what has been suggested on the basis of magnetisation and resistivity measurements [5,6], i.e. that conducting electrons participating in the double exchange do not interact with Cr$^{3+}$. As a consequence Cr$^{3+}$ ions are not magnetically ordered but their spins are randomly oriented within the $B$ sublattice.

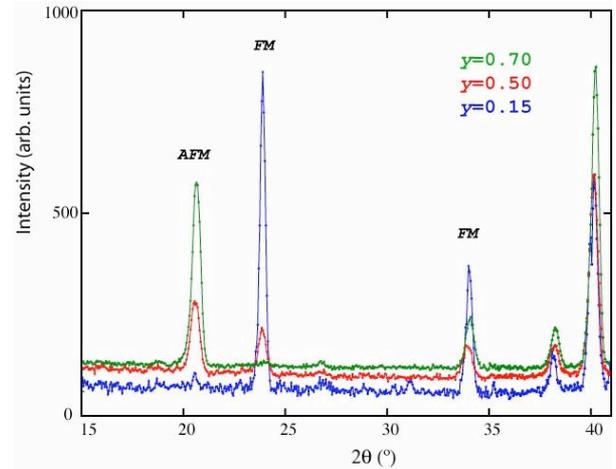

**Figure 5**: Low-angle portion of the NPD patterns of (La$_{0.70}$Ca$_{0.30}$)(Cr$_y$Mn$_{1-y}$)O$_3$ collected at 5 K. The intensity of the AFM peak progressively decreases moving from $y$ = 0.70 to 0.50, whereas it vanishes for the sample with $y$ = 0.15, which conversely shows only FM peaks (a weak nuclear contribution is present at 2θ~21.5°).

Unlike what observed in (La$_{0.70}$Ca$_{0.30}$)(Cr$_{0.5}$Mn$_{0.5}$)O$_3$, at this chromium concentration, the NPD data do show a decreases of SANS contribution with decreasing $T$. This is consistent with a percolation of FM clusters below the Curie temperature. At the same time incoherent scattering is partially suppressed on cooling and the background is decreased.

### 4. Conclusions

The nuclear and magnetic structures of (La$_{0.70}$Ca$_{0.30}$)(Cr$_y$Mn$_{1-y}$)O$_3$ samples ($y$ = 0.70, 0.50 and 0.15) at 290 K and 5 K have been investigated by means of neutron powder diffraction. For samples with $y$ = 0.70 and 0.50 an AFM structure ($G_x$-type) occurs in the Cr sub-lattice at 5 K, whereas the spins of the Mn ions are randomly oriented. The sample with $y$ = 0.15 exhibits FM ordering of the Mn-spins at 5 K whilst Cr is not involved in the double exchange. In conclusion no evidence for double or superexchange interactions among Cr$^{3+}$ and Mn cationic species has been observed. In particular it has been found that superexchange interactions among Cr$^{3+}$ ions prevail for $y$ = 0.70, 0.50, whereas for $y$ =

0.15 double exchange takes place among Mn cationic species only.

ACKNOWLEDGEMENTS
We acknowledge the School of Chemistry and the School of Physics and Astronomy of the University of Birmingham, the EPSRC (UK) and the EC (Training and Mobility of Researchers) for supporting part of this work. We thank L.Chapon for useful discussions during the International School of Neutron Scattering F. P. Ricci (Sardinia 2006).